\documentclass{Interspeech2024}
\usepackage{color,xcolor}
\usepackage{epsfig}
\usepackage{graphicx}
\usepackage{lipsum}
\usepackage{pifont}
\usepackage{array}
\usepackage{booktabs}
\usepackage{colortbl}
\usepackage{multirow}
\usepackage{float}
\usepackage{caption}
\usepackage{adjustbox}
\usepackage{subfigure}
\usepackage{footnote}
\usepackage{bbding}
\usepackage{cite}
\makesavenoteenv{tabular}
\makesavenoteenv{table}

\usepackage{amsmath,amsfonts,amssymb,bm}
\usepackage[super]{nth}

\usepackage{changepage}
\usepackage{extramarks}
\usepackage{lastpage}
\usepackage{setspace}
\usepackage{soul}
\usepackage{xspace}

\usepackage{url}
\usepackage{hyperref}
\hypersetup{colorlinks=True, urlcolor=black}

\usepackage{algorithm, algorithmic}
\usepackage{enumitem}
\usepackage{verbatim}
\usepackage{pifont}
\usepackage[acronyms]{glossaries}
\glsdisablehyper

\newcommand{\ignore}[1]{}

\makeatletter
\DeclareRobustCommand\onedot{\futurelet\@let@token\@onedot}
\def\@onedot{\ifx\@let@token.\else.\null\fi\xspace}

\makeatother

\definecolor{MyDarkBlue}{rgb}{0,0.08,1}
\definecolor{MyDarkGreen}{rgb}{0.02,0.6,0.02}
\definecolor{MyDarkRed}{rgb}{0.8,0.02,0.02}
\definecolor{MyDarkOrange}{rgb}{0.40,0.2,0.02}
\definecolor{MyPurple}{RGB}{111,0,255}
\definecolor{MyRed}{rgb}{1.0,0.0,0.0}
\definecolor{MyGold}{rgb}{0.75,0.6,0.12}
\definecolor{MyDarkgray}{rgb}{0.66, 0.66, 0.66}

\usepackage{pgfplots}
\DeclareUnicodeCharacter{2212}{−}
\usepgfplotslibrary{groupplots,dateplot}
\usetikzlibrary{patterns,shapes.arrows}
\pgfplotsset{compat=newest}

\interspeechcameraready

\title{LoRA-Whisper: Parameter-Efficient and Extensible Multilingual ASR}

\name[affiliation={1}]{Zheshu}{Song}
\name[affiliation={1}]{Jianheng}{Zhuo}
\name[affiliation={1}]{Yifan}{Yang}
\name[affiliation={1}]{Ziyang}{Ma}
\name[affiliation={2}]{Shixiong}{Zhang}
\name[affiliation={1,\dagger}]{Xie}{Chen}

\address{
  $^1$MoE Key Lab of Artificial Intelligence, AI Institute \\
  X-LANCE Lab, Department of Computer Science and Engineering, \\
  Shanghai Jiao Tong University, Shanghai, China \\
  $^2$Tencent AI Lab, USA }
\email{\{songzheshu, chenxie95\}@sjtu.edu.cn}

\newcommand\blfootnote[1]{
  \begingroup
  \renewcommand\thefootnote{}\footnote{#1}
  \addtocounter{footnote}{-1}
  \endgroup
}

\begin{document}

\maketitle


\begin{abstract}
Recent years have witnessed significant progress in multilingual automatic speech recognition (ASR), driven by the emergence of end-to-end (E2E) models and the scaling of multilingual datasets.
Despite that, two main challenges persist in multilingual ASR: language interference and the incorporation of new languages without degrading the performance of the existing ones.
This paper proposes LoRA-Whisper, which incorporates LoRA matrix into Whisper for multilingual ASR, effectively mitigating language interference.
Furthermore, by leveraging LoRA and the similarities between languages, we can achieve better performance on new languages while upholding consistent performance on original ones.
Experiments on a real-world task across eight languages demonstrate that our proposed LoRA-Whisper yields a relative gain of 18.5\% and 23.0\% over the baseline system for multilingual ASR and language expansion respectively.
\end{abstract}
\blfootnote{$\dagger$ Corresponding author}
\section{Introduction}
Automatic speech recognition (ASR) has traditionally concentrated on transcribing speech into written text for single languages \cite{CTC, RNN-T, AED, joint-ctc-aed,review}. Nevertheless, as the demand for cross-lingual communication grows and vast multilingual datasets \cite{MLS, fleurs, voxpopuli, commonvoice} become more accessible, attention has recently turned towards the development of massively multilingual ASR models.
With the emergence of large-scale multilingual speech recognition models such as Whisper \cite{Whisper}, Google USM \cite{USM}, and MMS \cite{MMS}, individuals now have the opportunity to construct \textbf{customized multilingual speech recognition models} tailored to specific languages based on these foundational models.

However, two significant challenges are still yet to be addressed in multilingual ASR. One is language interference, primarily stemming from language overlap, data imbalance, dialectal accents, etc. Another challenge involves incorporating new languages without compromising the performance of existing ones.
To resolve the former problem, there are a series of previous works attempting to mitigate this issue by leveraging language ID information \cite{LID-1, LID-2} or designing language-specific modules \cite{interference-cmm,interference-gated-language,interference-LRMoE,interference-informed-expert,interference-attention,interference-prompt,interference-A2} such as languages-specific encoders to differentiate each language. Besides, some works \cite{interference-pruning,interference-pruning2,interference-pruning3} utilize a pruning strategy in multilingual ASR with a dedicated sub-model for each language, while others propose new sampling methods \cite{interference-data-imbalance} to address the data imbalance issue. Although the methods mentioned above alleviate language interference to some extent, they are somewhat cumbersome in design and fail to account for language expansion. 
When new languages need to be integrated into a multilingual ASR system, a naive approach is to fine-tune the ASR model using data from these new languages. Unfortunately, this often results in \textit{catastrophic forgetting}, referring to the phenomenon that the recognition performance of base languages tends to decline. To solve the above problem, Li et al. \cite{expansion-lifelong} proposes lifelong learning \cite{lifelong-learning} solution which remedies the language interference problem by mixing base language data and new language data. However, this approach is inefficient and time-consuming. Libera et al. \cite{expansion-CLMASR} explores various continual learning methods \cite{ER,A-GEM,DER,Piggyback,EWC,LWF} to address the issue of catastrophic forgetting. While these approaches have helped alleviate the problem, it still persists.

Towards this end, we introduce LoRA-Whisper, a parameter-efficient and extensible model for multilingual ASR. LoRA \cite{lora}, originally introduced in natural language processing (NLP), effectively customizes large language models (LLMs) for specific domains. Drawing inspiration from this, it can also be used to tailor speech recognition models for specific languages. In practice, we assign a language-specific LoRA matrix for each language. This approach allows shared information across languages to be stored within the Whisper model, while language-specific information can be captured in the respective LoRA matrices. When incorporating a new language, a new LoRA matrix is assigned for it,  ensuring no impact on the performance of existing languages. Furthermore, by capitalizing on the similarities between the new language and base languages, we can enhance performance on the new language through improved initialization of the new LoRA matrix or by employing mixture of experts (MoE) \cite{moe}. Note that the foundational model is not restricted to Whisper but can encompass other open-source speech recognition models as well, we are simply utilizing Whisper as an exemplar in this paper. In summary, the contributions of this paper are as follows:
\begin{itemize}[leftmargin=0.8cm]
    \item We propose LoRA-Whisper to mitigate language interference and avoid catastrophic forgetting when incorporating new languages by attaching language-specific LoRA modules to the Whisper model.
    \item By utilizing the similarity between languages, notable performance improvement can be achieved on new languages via better initialization of the new LoRA matrix or the employment of MoE.
\end{itemize}

\section{Background}

\subsection{Whisper}
Whisper \cite{Whisper} is an encoder-decoder Transformer model that is capable of multiple speech tasks, including multilingual speech recognition, speech translation, language identification, and voice activity detection.
The input to Whisper is an 80-dimensional log-Mel spectrogram of 30 seconds length $\boldsymbol{X}=[\boldsymbol{x}_1,\boldsymbol{x}_2,\cdots,\boldsymbol{x}_T]$ where T denotes the context length. The encoder blocks encode the input speech feature into hidden representations $\boldsymbol{H}$ and the decoder blocks decode the hidden representations into text tokens $\boldsymbol{\hat{y}}$ recursively conditioned on previous tokens and special prompts $\boldsymbol{p}$. In formal terms, this process can be illustrated as follows:
\begin{equation}
    \boldsymbol{H} = AudioEncoder(\boldsymbol{X})
\end{equation}
\begin{equation}
    \hat{y_{t}}=TextDecoder\left(p, \hat{y}_{1: t-1}, \boldsymbol{H}\right)
\end{equation}

\subsection{LoRA}
LoRA \cite{lora} was initially introduced in the field of natural language processing (NLP) as a means to effectively tailor large language models (LLMs) for specific domains or downstream tasks. It was observed that the weights of pre-trained LLMs tend to exist mainly in a low-dimensional space. Taking inspiration from this observation, LoRA reduces the number of trainable parameters by learning pairs of rank decomposition matrices while keeping the original weights fixed.
Specifically, consider $i$-th feed forward layer $f_{i}(\boldsymbol{x})=\boldsymbol{W}_{i}\boldsymbol{x}+\boldsymbol{b}_{i}$, where $\boldsymbol{W}_{i} \in \mathbb{R}^{d_{1}\times d_{2}}$ and $\boldsymbol{b}_{i} \in \mathbb{R}^{d_{1}}$ denotes the frozen weight and bias. By applying LoRA, the forward process is modified as:
\begin{equation}
    f_{i}(\boldsymbol{x})=(\boldsymbol{W}_{i}+\Delta\boldsymbol{W}_{i})\boldsymbol{x}+\boldsymbol{b}_{i};\Delta\boldsymbol{W}_{i}=\boldsymbol{B}_{i}\boldsymbol{A_{i}}
\end{equation}
where $\boldsymbol{B}_{i} \in \mathbb{R}^{d_{1} \times r}$  and$\boldsymbol{A}_{i} \in \mathbb{R}^{r \times d_{2}}$ are the two trainable low-rank matrices, with the rank $r \ll min({d_{1},d_{2}})$.

\section{Methods}

In this paper, we mainly focus on tackling two challenges in multilingual ASR: one is language interference and the other is new language incorporation. Section 3.1 briefly outlines the main issues to be addressed in this paper. Section 3.2 and Section 3.3 introduce the methods used in multilingual ASR and language expansion in detail.

\subsection{Problem statement}
In our research, $n$ base languages are employed for the multilingual ASR experiment, alongside an additional $m$ new languages for the language expansion experiment, which can be denoted as 
$S_{1}=\{(\boldsymbol{X}_{i},\boldsymbol{Y}_{i}), i \in (1,n)\}$ and $S_{2}=\{(\boldsymbol{X}_{j},\boldsymbol{Y}_{j}), j \in (n+1,n+m)\}$ where $\boldsymbol{X}_{i}, \boldsymbol{Y}_{i}$ denotes the speech and transcription of $i$-th language.

In the multilingual ASR experiment, the aim is to alleviate language interference in $S_{1}$ and improve the performance of base languages.
The goal of language expansion is to incorporate $S_{2}$ into the multilingual model while maintaining the performance of $S_{1}$ unaffected, and leverage similarities between $S_{1}$ and $S_{2}$ to enhance the performance specifically on $S_{2}$.

\subsection{Multilingual ASR}

\begin{figure}[ht]
    \centering
    \includegraphics[scale=0.3]{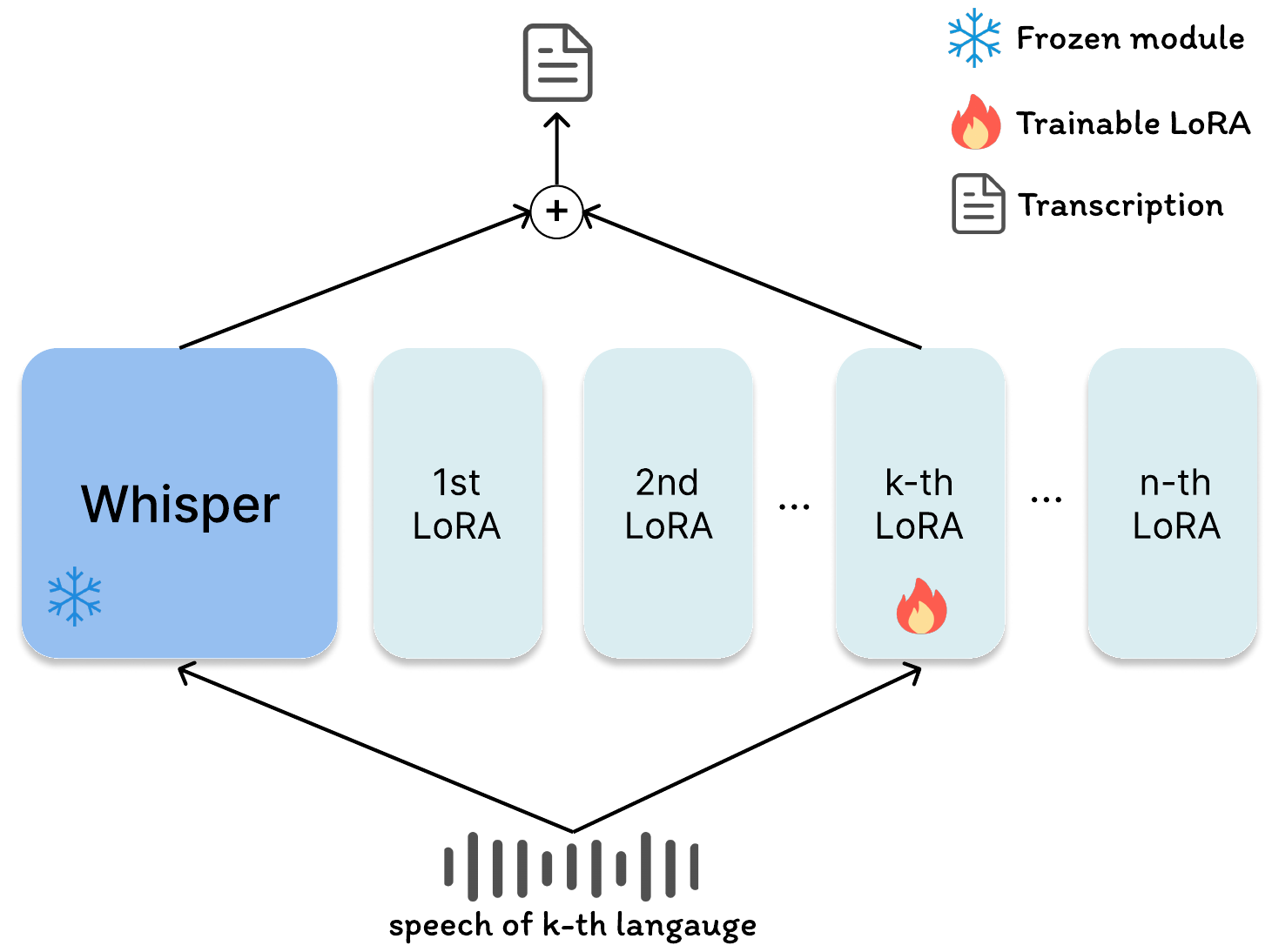}
    \caption{Architecture of LoRA-Whisper in multilingual ASR}
    \label{fig:LoRA-Whisper}
\end{figure}

Applying LoRA in multilingual ASR is an effective approach to mitigate language interference as shown in Figure \ref{fig:LoRA-Whisper}. For each language, a language-specific LoRA matrix is appended to the encoder and decoder of Whisper. When the input is a piece of speech of $k$-th language, it will activate $k$-th LoRA module and pass through Whisper and the corresponding LoRA module in the forward pass.

Under the LoRA-Whisper model, shared information across languages resides within the original Whisper model, while language-specific information is stored in the respective LoRA module. As a result, not only is the language interference problem skillfully avoided, but the performance of Whisper on specific languages is also significantly enhanced.
\subsection{Language expansion}

\begin{figure*}[t]
    \centering
    \includegraphics[scale=0.43]{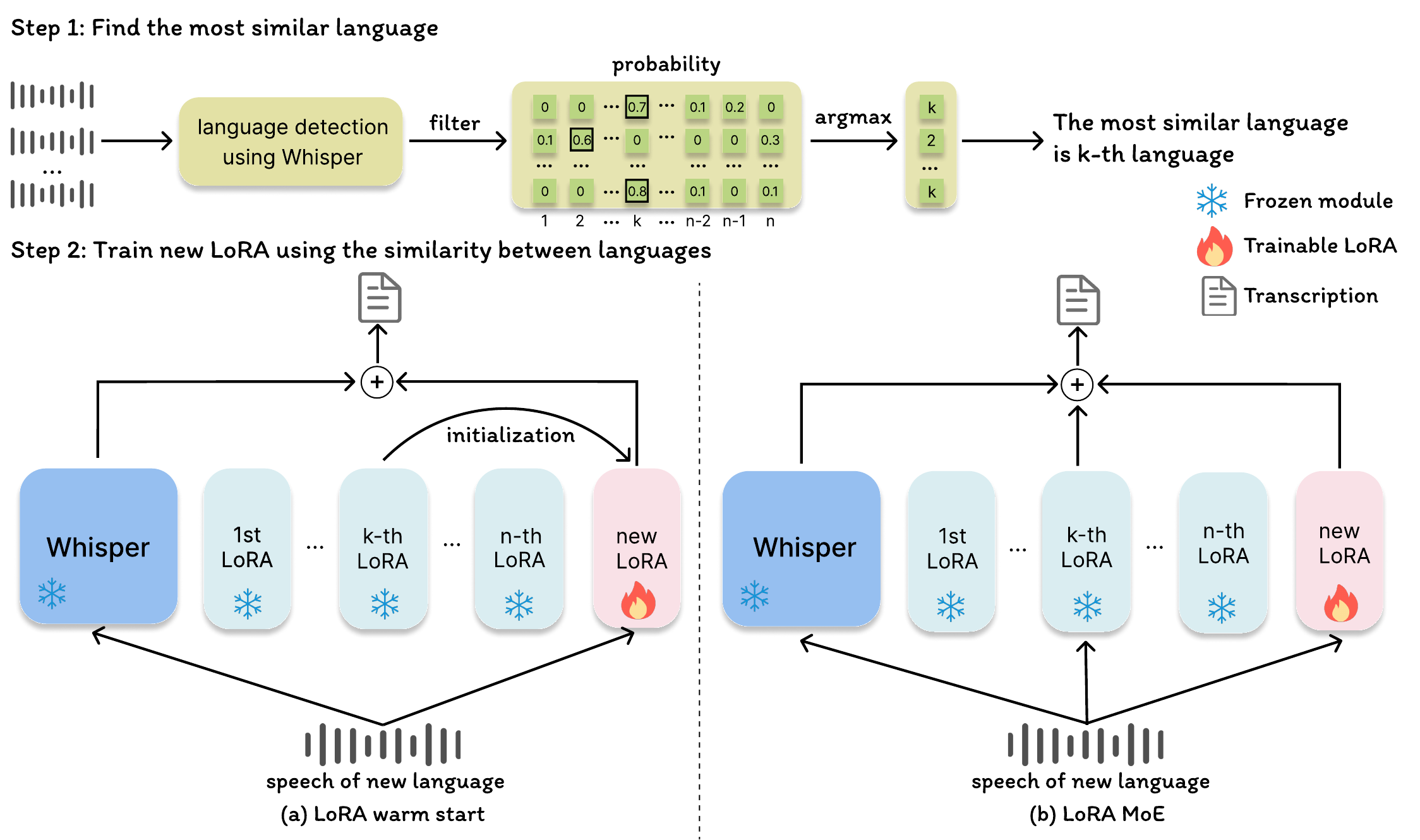}
    \caption{Architecture of LoRA-Whisper in language expansion. Left: LoRA warm start, Right: LoRA MoE}
    \label{fig:expansion}
\end{figure*}

Apart from mitigating language interference, LoRA can also be naturally extended for language expansion, preventing catastrophic forgetting.
Moreover, harnessing the similarities across languages can facilitate more effective training for new languages. Consequently, we introduce two effective methods for language expansion, namely LoRA warm start and LoRA MoE as depicted in Figure \ref{fig:expansion}. These methods involve two steps.

\textbf{Step 1: Find the most similar language} When incorporating a new language into the existing model, we first randomly sample $M$ audio segments from the new language data. These audio are then processed using the Whisper model for language detection. The output provides a probability distribution over all languages.
In our experiment, the focus lies solely on the languages employed in the aforementioned multilingual ASR experiment. Specifically, we extract the probabilities associated with these languages and normalize them, which can be denoted as $\boldsymbol{p_{i}} = [p_{i1}, p_{i2}, \cdots, p_{in}]; i=1, \cdots, M$. When incorporating a new language, the most similar language to it can be found by measuring the similarity between the new language and base languages, which is defined as follows:
\begin{equation}
    sim_{k}=\frac{\sum_{i=1}^{M} \mathbb{I}(k=\mathop{\arg\max}\limits_{j}p_{ij})}{M} \quad \text { for } k=1, \cdots, n
\end{equation}
where $\mathbb{I}$ is indicator function and $sim_{k}$ is the defined similarity between new language and $k$-th base language.

\textbf{Step 2: Continual training on new languages} After finding the most similar one, we can leverage the information from base languages to facilitate the training of the new language. In LoRA warm start, the new LoRA matrix is initialized from the LoRA matrix of its most similar language. In LoRA MoE, two LoRA modules are selected in the forward pass to assit in the training of the new language.

In general, aside from avoiding the issue of catastrophic forgetting, LoRA-Whisper enables the exploitation of similarities between new language and base languages, leading to better performance on the new language.
\section{Experiments}

\subsection{Dataset}
 Our experiments are conducted on MLS dataset\cite{MLS} and FLEURS\cite{fleurs} dataset as shown in Table \ref{data}. Due to limited resources, we only focus on Polish, Portuguese and Italian in MLS dataset. Similarly, five languages are selected from FLEURS dataset, namely Chinese, Danish, Greek, Welsh and Japanese. Four languages are employed in the multilingual ASR experiment and the remaining four languages are used as new languages in the language expansion experiment.
 
\begin{table}[h]
\renewcommand\arraystretch{1.2}
\caption{Statistics of training and testing data (in hours)}
\label{data}
\begin{tabular}{c|c|cc}
\toprule[1.05pt]
\textbf{Experiment}                          & \textbf{Language}       & \textbf{Train} & \textbf{Test} \\ \hline
\multirow{4}{*}{Multilingual ASR}   & Polish(PL)     & 103.0   & 2.0    \\
                                    & Portuguese(PT) & 160.0   & 5.0    \\
                                    & Italian(IT)    & 247.0   & 5.0    \\
                                    & Chinese(ZH)    & 9.7    & 3.1    \\ \hline
\multirow{4}{*}{Language expansion} & Danish(DA)     & 7.5   & 2.9    \\
                                    & Greek(EL)      & 10.0    & 1.9  \\
                                    & Welsh(CY)      & 12.2    & 4.3  \\
                                    & Japanese(JA)   & 7.4   & 2.4  \\
\bottomrule[1.05pt]
\end{tabular}
\end{table}
\vspace{-0.3cm}
\begin{table}[ht]
\caption{WER on MLS Polish under different LoRA rank \textit{r}}
\label{different_rank}
\tabcolsep=0.15cm
\renewcommand\arraystretch{1.1}
\begin{tabular}{c|ccccc}
\toprule[1.05pt]
Rank \textit{r}           & 8    & 16   & 32            & 48    & 64    \\ \hline
\#LoRA param & 3.2M & 6.5M & 13.0M         & 19.5M & 26.0M \\ \hline
WER   & 9.00 & 8.54 & \textbf{7.94} & 8.00  & 8.00   \\ 
\bottomrule[1.05pt]
\end{tabular}
\end{table}
\begin{table*}[t]
\caption{Comparison of Whisper-small and LoRA-Whisper in WER/CER for base languages.}
\vspace{-1.0mm}
\label{exp1}
\tabcolsep=0.376cm
\renewcommand\arraystretch{1.287}
\begin{tabular}{c|c|c|c|ccccc}
\toprule[1.05pt]
\textbf{ID} & \textbf{Model} &  \textbf{Finetune} & \textbf{\#Train param} & \textbf{PL} & \textbf{PT} & \textbf{IT} & \textbf{ZH} & \textbf{Avg} \\ \hline
E1 & \multirow{3}{*}{Whisper-small} & No & - & 10.90 & 13.82 & 20.43 & 9.25 & 13.60 \\
E2 &     & Full(multilingual) & 240M & 8.30 & 13.34 & 10.69 & 14.37 & 11.68 \\
E3 &     & Full(monolingual) & 240M*4 & \textbf{7.65} & \textbf{10.21} & \textbf{10.27} & \textbf{8.64} & \textbf{9.19} \\ \hline
E4 & LoRA-Whisper   & LoRA & 13M*4 & 7.94 & 10.81 & 10.49 & 8.82 &  9.52 \\
\bottomrule[1.05pt]
\end{tabular}
\end{table*}
\begin{table*}[t]
\renewcommand\arraystretch{1.287}
\tabcolsep=0.138cm
\caption{Experimental results of language expansion. E1 denotes the results of original Whisper model and E2 denotes the results on base languages before language expansion. Full means multilingual full fine-tune with new language data and full+ means using both new language and base language data. Seed model serves as an initialization for continual training.}
\vspace{-1.0mm}
\label{exp2}
\begin{tabular}{c|c|c|c|c|ccccc|ccccc}
\toprule[1.05pt]
\multirow{2}{*}{\textbf{ID}} & \multirow{2}{*}{\textbf{Model}} & \multirow{2}{*}{\textbf{Finetune}} & \multirow{2}{*}{\begin{tabular}[c]{@{}c@{}}\textbf{Seed} \\ \textbf{model}\end{tabular}} & \multirow{2}{*}{\begin{tabular}[c]{@{}c@{}}\textbf{\#Train} \\ \textbf{param}\end{tabular}} & \multicolumn{5}{c|}{\textbf{New languages}}    & \multicolumn{5}{c}{\textbf{Base languages}} \\ \cline{6-15} 
                              &   &   &  &  & \textbf{DA}  & \textbf{EL}  & \textbf{CY}  & \textbf{JA} & \textbf{Avg} & \textbf{PL} & \textbf{PT} & \textbf{IT} & \textbf{ZH} & \textbf{Avg}  \\ \hline
E1 & \multirow{4}{*}{Whisper-small}   & No & E1 & -   & 33.97 & 31.81 & 58.62 & 12.04 & 34.11 & 10.90 & 13.82 & 20.43 & 9.25 &  13.60  \\  
E2 &                               & No & E2 &  - & - & - & - & - & - & 8.30 & 13.34 & 10.69 & 14.37 &  11.68 \\
E5 &                               & Full & E2 & 240M  & 38.88 & 26.08  & 32.90 & 15.46  & 28.33 & 17.25 & 25.56 & 17.41  & 68.93 & 32.29 \\  
E6 &                               & Full+ & E2 & 240M & 41.35 & 28.89  & 33.65 & 16.61  & 30.13 & 9.77  & 15.41 & 12.16  & 16.07  & 13.35  \\ \hline
E7 & \multirow{3}{*}{LoRA-Whisper}  & LoRA & E4 & 13M*4  & 28.28 & 22.84 & 30.63 & 10.11 & 22.97 & \textbf{7.94} & \textbf{10.81} & \textbf{10.49} & \textbf{8.82} & \textbf{9.52} \\
E8 &                               & Warm start & E4 & 13M*4 & \textbf{27.45} & 21.77 & \textbf{28.05} & \textbf{10.03}  & 21.83  & \textbf{7.94} & \textbf{10.81} & \textbf{10.49} & \textbf{8.82} & \textbf{9.52} \\ 
E9 &                               & LoRA MoE & E4 & 13M*4 & 27.56 & \textbf{21.56} & 28.07 & \textbf{10.03} & \textbf{21.81} & \textbf{7.94} & \textbf{10.81} & \textbf{10.49} & \textbf{8.82} & \textbf{9.52} \\
\bottomrule[1.05pt]
\end{tabular}
\end{table*}

\subsection{Training configuration}
We evaluate the performance of our proposed method on Whisper-small. The impact of different LoRA rank on model performance is studied as shown in Table \ref{different_rank}. It can be seen that best performance is achieved under $r=32$. Hence, low-rank matrices where rank $r=32$ are added to the attention layer ${\{\boldsymbol{W}_{k},\boldsymbol{W}_{q},\boldsymbol{W}_{v}\}}$ and fully-connected layer $\boldsymbol{W}_{fc}$ in each transformer layer in both encoder and decoder.

In the training stage, we fix all the parameters of Whisper and optimize the language-specific LoRA modules with AdamW \cite{adamw} with a peak learning rate of 1e-4. The number of training epochs is set to 10. All models are trained with 2 NVIDIA RTX 3090 24GB GPUs.
In the testing stage, beam search with $beamsize=5$ is employed to decode the test set.

\subsection{Results and analysis}

\textbf{Multilingual ASR} The results of multilingual ASR are summarized in Table \ref{exp1}. It can be observed that monolingual full fine-tuning yields optimal outcomes at the cost of training and maintaining four systems, leading to higher maintenance costs. In contrast to monolingual full fine-tuning, multilingual full fine-tuning would lead to significant language interference issues when all training data are mixed together, with the average WER increasing from 9.19\% to 11.68\% (Table \ref{exp1}: E2 \textit{vs} E3). Our proposed approach, LoRA-Whisper, effectively eliminates interference between languages by incorporating language-specific LoRA matrices. In comparison to multilingual full fine-tuning, LoRA-Whisper yields superior results with fewer trainable parameters (Table \ref{exp1}: E2 \textit{vs} E4). Furthermore, when compared to monolingual full fine-tuning, it achieves performance almost on par while requiring training of only 5\% of parameters (Table \ref{exp1}: E3 \textit{vs} E4).

\textbf{Language expansion}
As can be seen from Table \ref{exp2}, when adding new languages, fine-tuning the model solely with new language data results in serious catastrophic forgetting, with WER on base languages nearly triples compared to the previous result (Table \ref{exp2}: E2 \textit{vs} E5). A simple way to mitigate catastrophic forgetting is to mix a portion of original data with new data, and then train the model on this combined dataset, as indicated by E6 outlined in Table \ref{exp2}. For each base language, we extract 5 hours of training data and merge them with new language data. It can be seen that this can alleviate the phenomenon of catastrophic forgetting to a certain degree but at the sacrifice of performance on new languages caused by language interference. The LoRA-Whisper we propose can solve this problem in an elegant way. Without affecting the performance of base languages, the LoRA-Whisper model can make use of the similarity between languages to flexibly expand new languages. 
Our proposed methods (LoRA warm start and LoRA MoE) yield a relative gain of 23\% and 5\% over full fine-tuning and LoRA fine-tuning respectively, demonstrating the efficacy of our model (Table \ref{exp2}: E8 \textit{vs} E5 \& E7, E9 \textit{vs} E5 \& E7).

\subsection{Ablation study}
In language expansion, the most similar language to the new language are chosen to assist in training. To validate the effectiveness of this approach, a series of experiments have been conducted on LoRA warm start.

Utilizing Step 1 as illustrated in Figure \ref{fig:expansion}, we can obtain that Danish and Greek are most similar to Portuguese, while Welsh and Japanese are most similar to Polish and Chinese respectively. From Table \ref{similarity}, it is evident that the model attains optimal performance on new languages when the new LoRA matrix is initialized from the LoRA matrix of the most similar language. Another noteworthy observation is that initializing it with less relevant language's LoRA matrix may result in decreased model performance compared to training from scratch, underscoring the significance of selecting the most similar language to the new one.
\vspace{-1.0mm}
\begin{table}[h]
\caption{Ablation study on LoRA warm start. - means the new LoRA matrix is trained from scratch.}
\vspace{-1.0mm}
\label{similarity}
\renewcommand\arraystretch{1.265}
\tabcolsep=0.1cm
\begin{tabular}{c|c|ccccc}
\toprule[1.05pt]
\multirow{2}{*}{\begin{tabular}[c]{@{}c@{}}\textbf{New} \\ \textbf{language}\end{tabular}} & \multirow{2}{*}{\begin{tabular}[c]{@{}c@{}}\textbf{Most similar}\\ \textbf{language}\end{tabular}} & \multicolumn{5}{c}{\textbf{Initialization}}  \\ \cline{3-7} 
                                                                          &                                                                                  &  \textbf{-} & \multicolumn{1}{c}{\textbf{PL}}  & \multicolumn{1}{c}{\textbf{PT}}  & \multicolumn{1}{c}{\textbf{IT}} & \textbf{ZH} \\ \hline
DA                                                                    & PT                                                                       & 28.28 & \multicolumn{1}{c}{30.19}          & \multicolumn{1}{c}{\textbf{27.45}} & \multicolumn{1}{c}{28.26}   & 28.15          \\ 
EL                                                                     & PT                                                                       & 22.84 & \multicolumn{1}{c}{23.1}           & \multicolumn{1}{c}{\textbf{21.77}} & \multicolumn{1}{c}{21.81}   & 22.19          \\ 
CY                                                                     & PL                                                                           & 30.63 & \multicolumn{1}{c}{\textbf{28.05}} & \multicolumn{1}{c}{28.42}          & \multicolumn{1}{c}{28.5}    & 29.34       \\ 
JA                                                                  & ZH                                                                          & 10.11 & \multicolumn{1}{c}{11.69}          & \multicolumn{1}{c}{10.34}          & \multicolumn{1}{c}{10.36}   & \textbf{10.03} \\ 
\bottomrule[1.05pt]
\end{tabular}
\end{table}

\vspace{-4.0mm}
\subsection{Limitation}
The limitation of LoRA-Whisper lies in the model size will become larger as the number of languages continues to increase. Hence, future research will explore the sharing of LoRA within multiple similar languages and expand to more languages.

\section{Conclusions}
In this study, we introduce LoRA-Whisper, a parameter-efficient and extensible multilingual ASR model. By attaching language-specific LoRA modules to the Whisper model, our approach effectively solves the problem of language interference and achieves better language expansion via LoRA warm start or LoRA MoE, allowing people build customized multilingual speech recognition models based on these speech foundation models. We hope that our study can facilitate the research on language expansion in multilingual ASR.
\section{Acknowledgements}
This work was supported by the National Natural Science Foundation of China  (No. 62206171 and No. U23B2018), Shanghai Municipal Science and Technology Major Project under Grant 2021SHZDZX0102 and the International Cooperation Project of PCL and Tencent AI Lab Rhino-Bird Focused Research Program.

\bibliographystyle{IEEEtran}
\bibliography{mybib}
\end{document}